\documentclass[final,5p,times,twocolumn]{elsarticle} 

\usepackage{lineno,hyperref}
\usepackage{subcaption}
\usepackage{amsmath}
\usepackage{amssymb}

\journal{Nuclear Instruments and Methods A}
\begin{document}

\begin{frontmatter}

\title{ALICE Fast Interaction Trigger Upgrade}
\author{Krystian Roslon\corref{mycorrespondingauthor}}
\cortext[mycorrespondingauthor]{Corresponding author}
\ead{krystian.roslon@pw.edu.pl}
\author{for the ALICE collaboration}
\address{Faculty of Physics, Warsaw University of Technology, \\ Koszykowa 75, 00-662 Warsaw, Poland}

\begin{abstract}
This proceeding provides an expanded overview of the Fast Interaction Trigger (FIT) system performance, focusing on new developments such as the prospective integration of the ALICE Low-Level Front-End Device (ALFRED) into the Detector Control System (DCS) and an upgraded Front-End Electronics (FEE) approach to enhance dynamic range and operational reliability. The first upgrade is dedicated to integrating FIT with ALICE central systems, while the second aims to improve signal processing from the scintillation arrays (FV0 and FDD). Additionally, we propose forward-detector applications in future ALICE upgrades (Run 5 and beyond).

We also present the latest performance results, illustrated with relevant plots, including collision-time measurements for pp and Pb--Pb collision systems, collision centrality determination based on the amplitude signals from the FT0 detector, trigger performance metrics, and the improved DCS architecture.
\end{abstract}


\end{frontmatter}


\section{Introduction}
\label{sec:intro}

A Large Ion Collider Experiment \cite{Aamodt:2008zz}(ALICE) focuses on investigating the physics of strongly interacting matter under conditions of extreme energy densities, with a particular emphasis on the characteristics of the Quark–Gluon Plasma. During the Long Shutdown 2 (LS2) of the LHC, the collaboration undertook several upgrades to improve tracking and vertexing and prepared the ALICE apparatus to be able to scope with 50 kHz interaction rate of Pb--Pb collisions and data taking with continuous readout. One of these upgrades is FIT \cite{Slupecki:2022fch}, which has been installed in the forward regions adjacent to the interaction point (IP).

FIT delivers low-latency ($<$425 ns) minimum-bias triggers at rates up to 1 MHz in pp collisions, monitors beam background and luminosity in real time, and provides precise collision-time measurements for the time-of-flight detector. In addition, it enables centrality and multiplicity determination, event-plane estimation for heavy-ion collisions, and detection of diffractive events.

\begin{figure}[!ht]
    \centering
    \includegraphics[width=.86\linewidth]{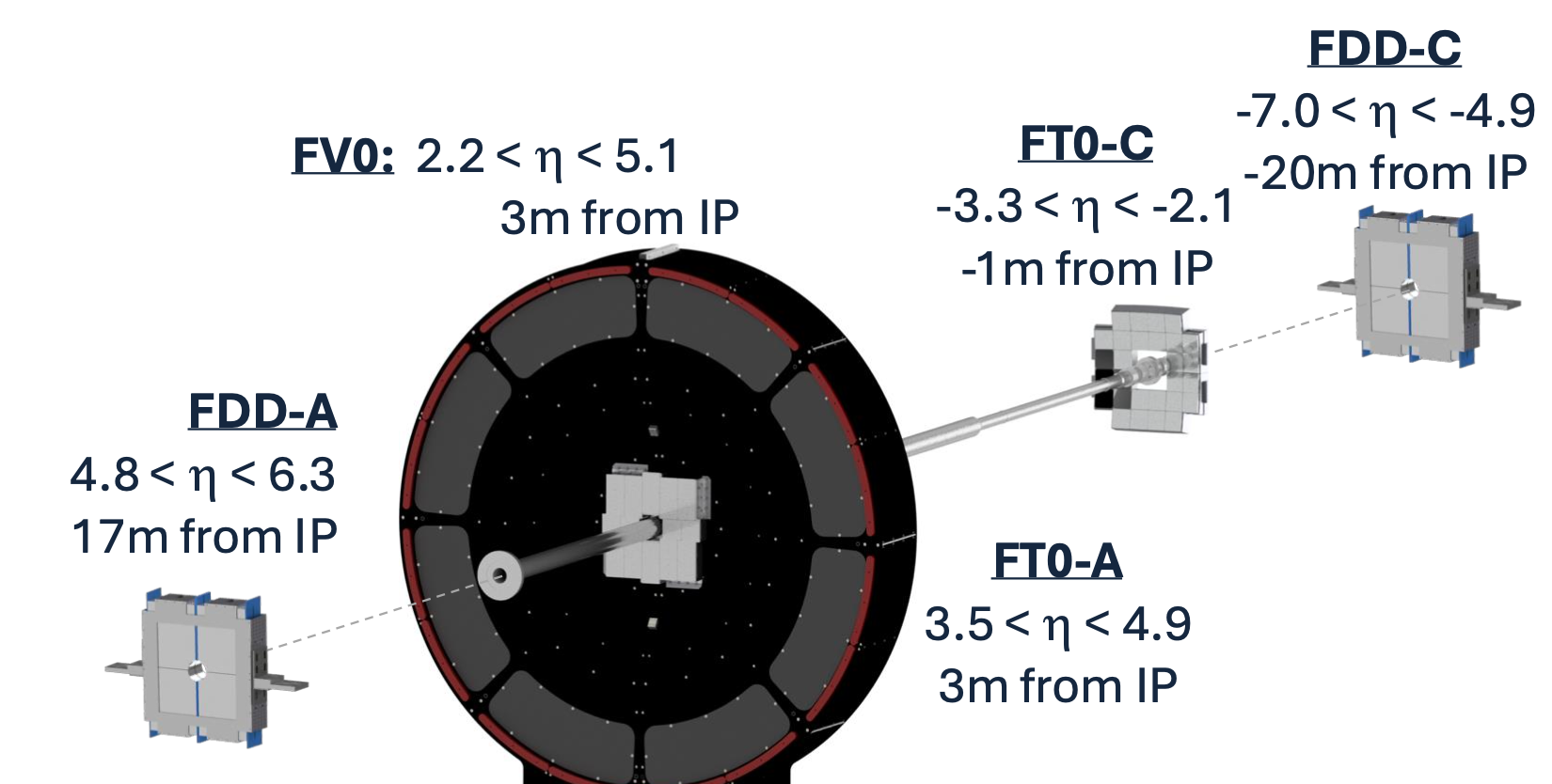}
    \caption{Layout of the FIT detector system, showing FT0, FV0, and FDD.}
    \label{fig:FIT}
\end{figure}

Figure~\ref{fig:FIT} illustrates the layout of the three main subdetectors within FIT: the Cherenkov-based FT0, and two scintillator-based detectors (FV0 and FDD), all positioned in different pseudorapidity regions.

Since installation, FIT has operated flawlessly, adhering strictly to the established plan without any maintenance issues. Current efforts are concentrated on two major upgrades: modifying the FEE and implementing a new architecture for the Detector Control System \cite{Roslon:2025tym}. These modifications aim to increase the dynamic range of the detectors and integrate the ALFRED framework \cite{Chochula:2018vfx}. Subsequently we anticipate exploring opportunities for innovative forward detectors in upcoming upgrades of the ALICE project.

\section{Operational Overview and the LHC Timeline}
\label{sec:LHC}

\begin{figure}[!ht]
    \centering
    \includegraphics[width=.9\linewidth]{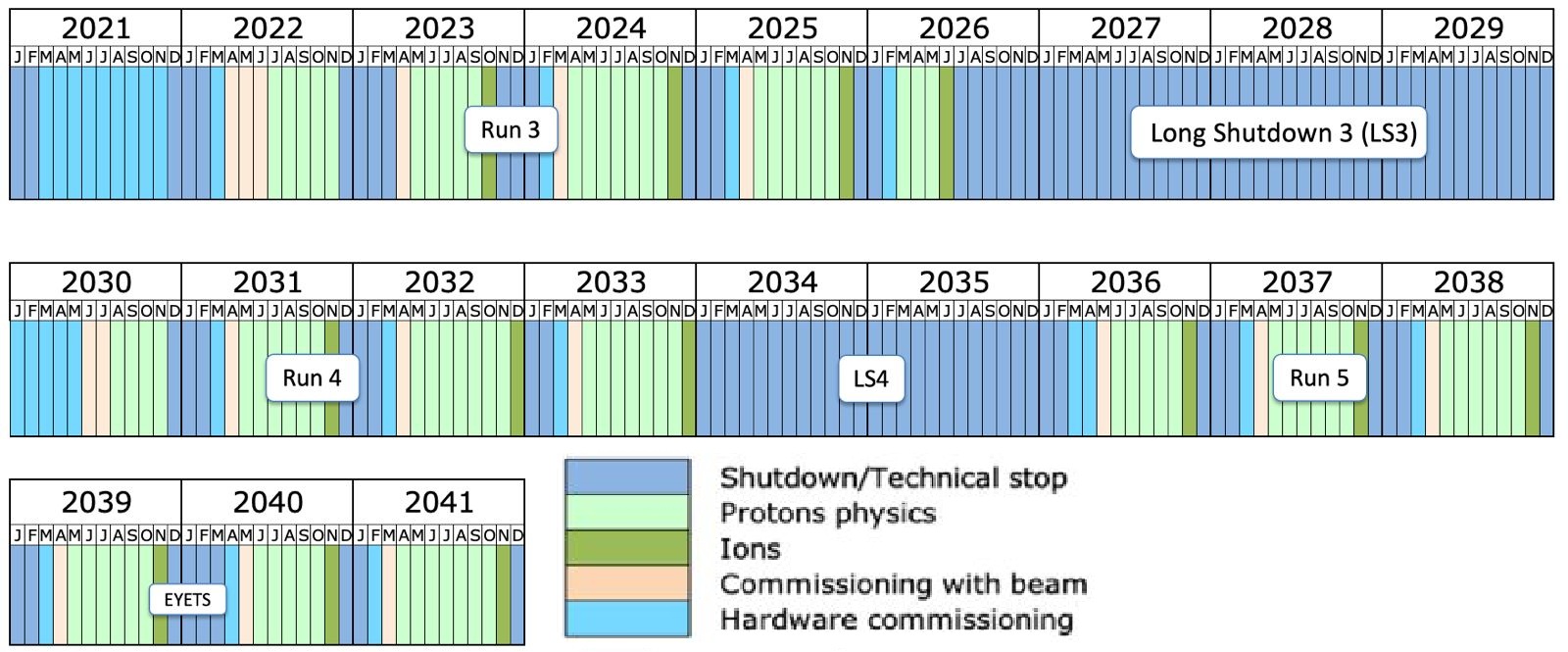}
    \caption{A mid-term LHC schedule (updated to September 2024).}
    \label{fig:LHC}
\end{figure}

The LHC aims to reach higher luminosities in proton--proton and heavy-ion collisions, with ongoing improvements in injector performance and beam strategies. Figure~\ref{fig:LHC} displays a longer-term schedule for the LHC, showing the shutdowns and key operational periods. ALICE has capitalised on these intervals to install and commission upgraded detector systems, including FIT.

\section{FIT Performance in Run 3}
\label{sec:performance}

\subsection{Subdetector Functions}

\textbf{FT0}: A pair of Cherenkov arrays (FT0-A and FT0-C), each composed of quartz radiators coupled to microchannel plate photomultiplier tubes (MCP-PMTs PLANACON\textsuperscript{\textregistered} XP85012 \cite{Melikyan:2020owp}). It is the fastest FIT subdetector, providing excellence time resolutions and minimal latency ($<$425 ns) for the online trigger.\\
\textbf{FV0}: A large scintillator ring (consisting of multiple segments arranged radially) that measures charged-particle multiplicities in forward rapidities. FV0 readings complement those of FT0 for centrality and event-plane analyses.\\
\textbf{FDD}: Two double-layered scintillator arrays located further along the beam pipe, which is used for very-forward diffractive event tagging and background monitoring.

\subsection{Collision-Time Resolution and Vertexing}

\begin{figure}[!ht]
\centering
\begin{subfigure}{.25\textwidth}
  \centering
  \includegraphics[width=.90\linewidth]{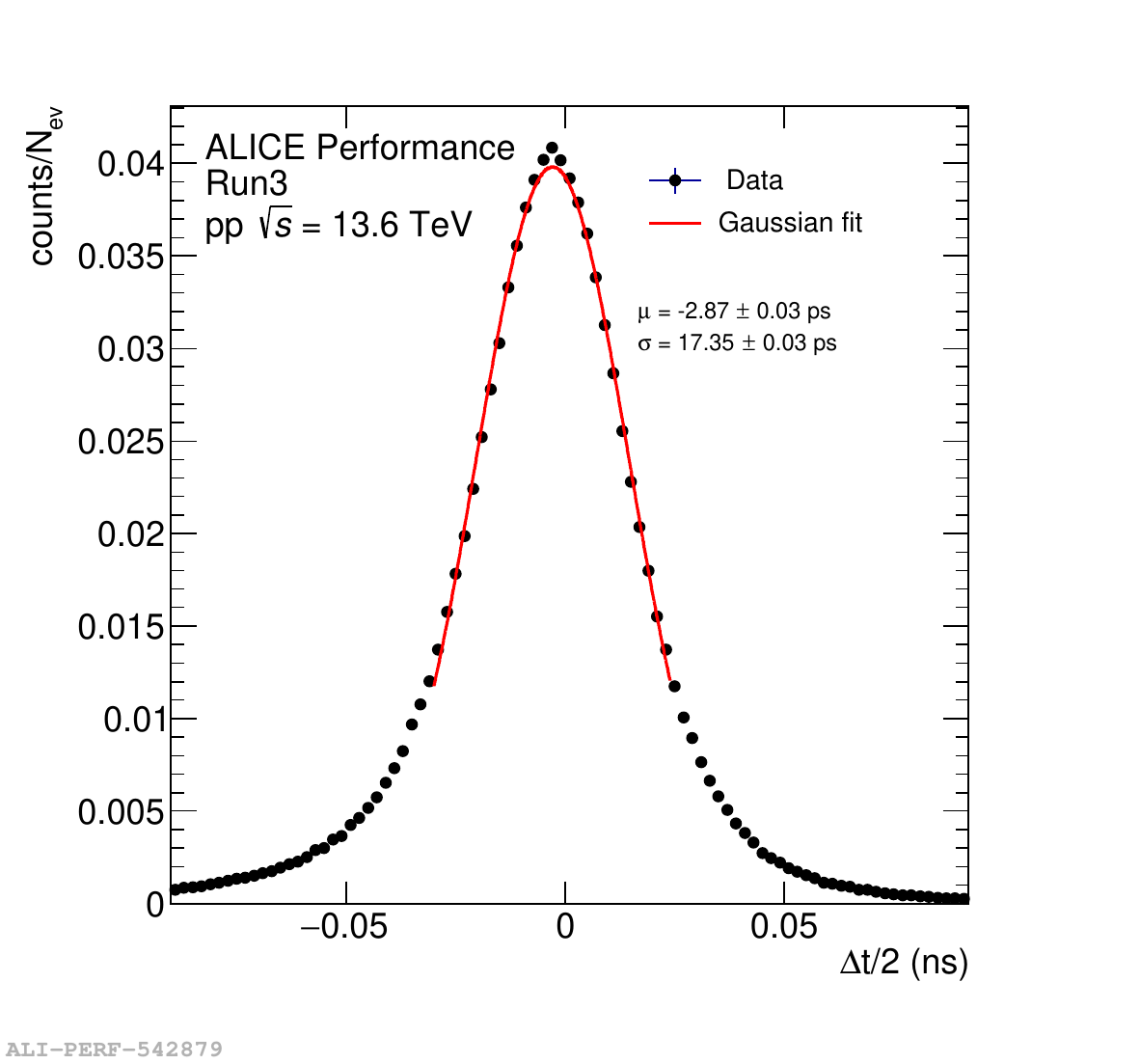}
  \caption{pp 13.6 TeV}
  \label{fig:FT0_Time_pp}
\end{subfigure}%
\begin{subfigure}{.25\textwidth}
  \centering
  \includegraphics[width=.9\linewidth]{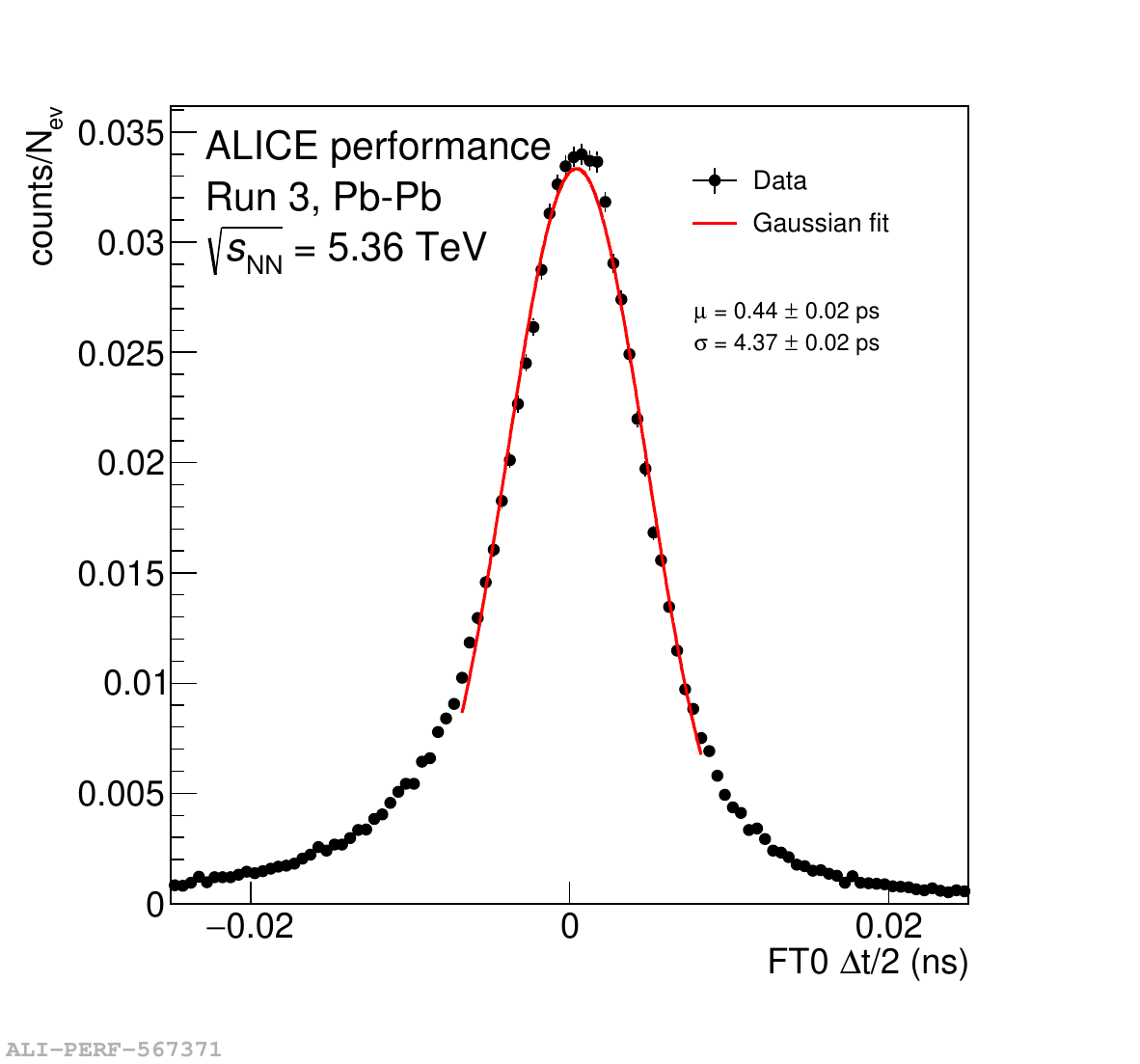}
  \caption{Pb--Pb 5.36 TeV}
  \label{fig:FT0_Time_PbPb}
\end{subfigure}
\caption{FT0 timing resolutions in (a) pp and (b) Pb--Pb collisions.}
\label{fig:FT0_Time_resolution}
\end{figure}

Thanks to its Cherenkov-based design, the FT0 in Run~3 has proven to be the top performer in timing. In proton--proton collisions at $\sqrt{s}=13.6$\,TeV, the time resolution is about 17\,ps while in Pb--Pb collisions at $\sqrt{s_{\mathrm{NN}}}=5.36$\,TeV, benefiting from higher particle multiplicities, the FT0 achieves $\sim$4.4\,ps as shown in figure~\ref{fig:FT0_Time_resolution}.

Such precise timing also facilitates online vertex determination. Combining time measurements from both FT0 sides yields a fast and reliable vertex estimate, which has been cross-checked against offline tracking detectors and found to be in good agreement \cite{Melikyan:2024uja}.

\subsection{Centrality and Multiplicity Measurements}

\begin{figure}[!ht]
    \centering
    \includegraphics[width=.82\linewidth]{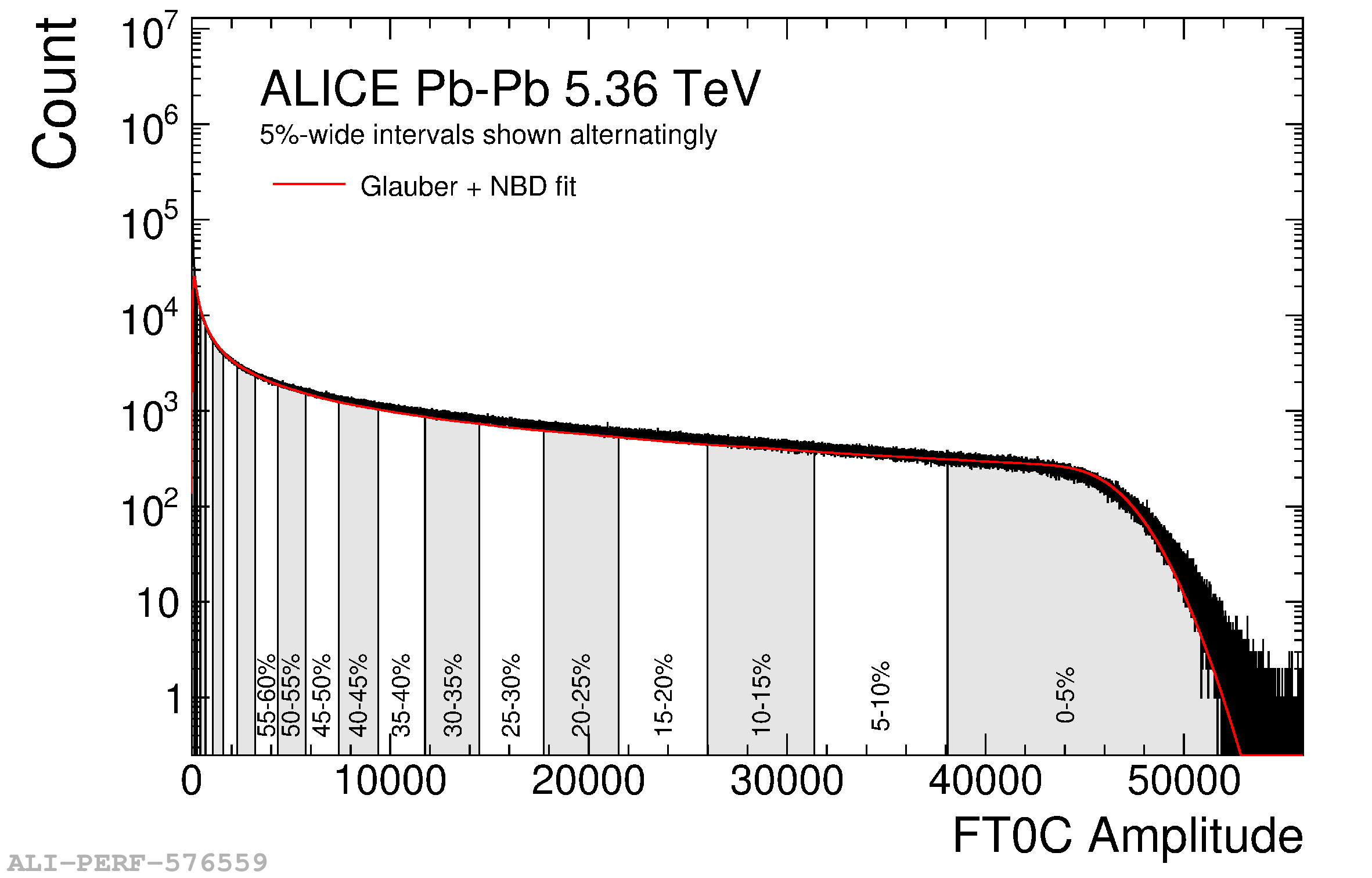}
    \caption{Illustrative centrality determination in Pb--Pb collisions at $\sqrt{s_{\rm NN}}=5.36$ TeV, derived from FT0 amplitude measurements.}
    \label{fig:Centrality}
\end{figure}

Beyond timing, FIT data is used for the determination of event centrality. Signals from FIT measure forward charged-particle flux, which, when compared to Glauber + NBD fit, provides centrality classes (e.g.\ 0--5\%, 5--10\%, etc.). Figure~\ref{fig:Centrality} depicts an example amplitude distribution in Pb--Pb at 5.36\,TeV based on the FT0 detector. FV0 also provides forward-multiplicity coverage at large pseudorapidities, complementing the FT0. Correlations of the scintillator signals from the FV0 with those from the FT0 help cross-check multiplicity-based triggers.

\subsection{FT0 Triggering in PbPb collisions}

\begin{figure}[!ht]
    \centering
    \includegraphics[width=.65\linewidth]{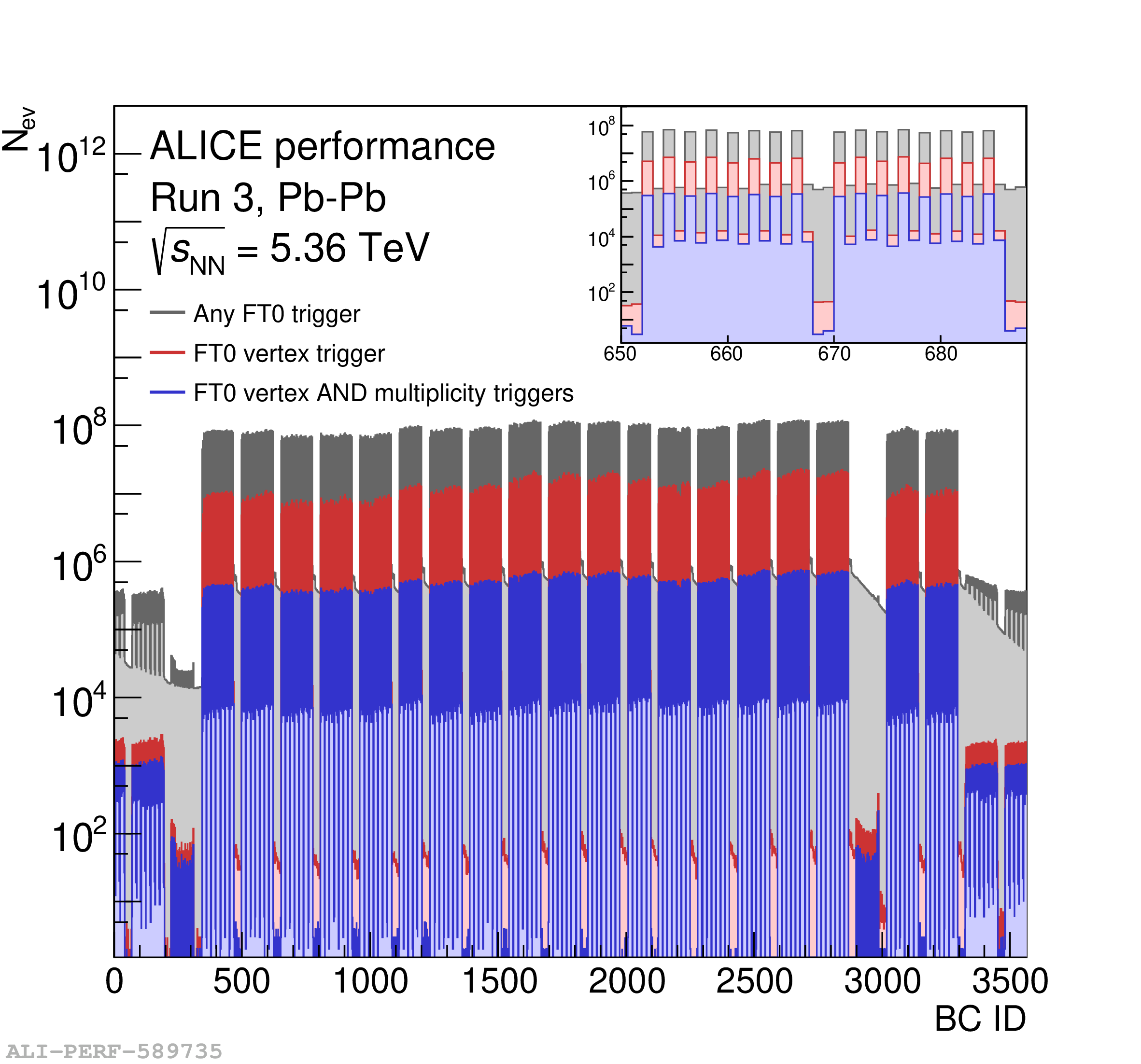}
    \caption{FT0 Bunch Crossing (BC) trigger counts distribution in Pb--Pb collisions at $\sqrt{s_{\rm NN}}=5.36$ TeV}
    \label{fig:Trigger}
\end{figure}

The FT0 subsystem provides several trigger signals—ORA, ORC, VTX, and amplitude-based triggers—specifically designed for efficient ion-collision selection. In particular, amplitude-based triggers like the semi-central (SC) and central (CE) rely on integrated charge from the FT0A and FT0C to discriminate event centrality in real time. The FT0 typically has about \(14\,\text{ADC channels}\) per minimum ionizing particle (MIP) and the SC threshold corresponds to ~ 35 MIPs and CE to ~ 1433 MIPs. These values are derived from the fit done in Fig.~\ref{fig:Centrality}. The VTX trigger, which requires a coincidence between the FT0-A and FT0-C, further enhances vertex-based event tagging. Figure~\ref{fig:Trigger} shows that these triggers collectively achieve clean event selection with minimal background contamination.

\subsection{Detector Control System}

\begin{figure}[!ht]
    \centering
    \includegraphics[width=.85\linewidth]{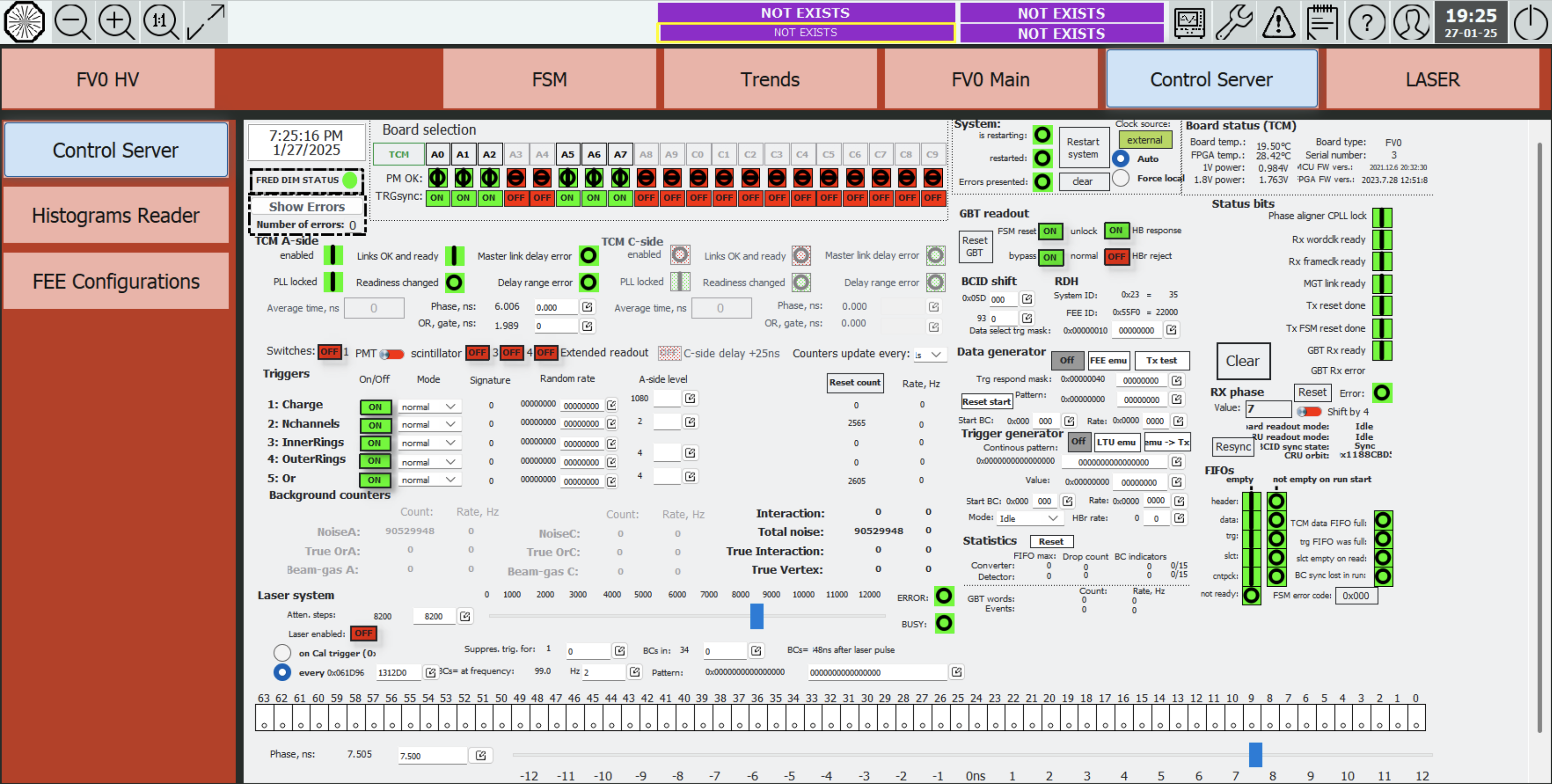}
    \caption{FV0 subdetector’s SCADA interface.}
    \label{fig:DCS}
\end{figure}

The Detector Control System for all FIT subdetectors has been extensively developed and unified, ensuring a high degree of automation and reliability. This consolidation includes a Final State Machine (FSM), which manages transitions between operational modes—such as off, standby, calibration, and data-taking with minimal user intervention. Key functionalities of the control server have undergone thorough testing, demonstrating stable performance under nominal and stress-test conditions. As a result, the number of on-call interventions required to maintain FIT operations has decreased markedly compared to previous years. This reduction stems from improved error handling, streamlined interfaces, and enhanced monitoring tools that allow potential issues to be identified and resolved proactively. Figure~\ref{fig:DCS} shows an example Supervisory Control And Data Acquisition (SCADA) interface for the FV0 subdetector. Processes like powering, threshold tuning, and monitoring of system stability are all coordinated through standardised interfaces.

\section{Detector Ageing and Upgrade Plans}
\label{sec:upgrade}

\subsection{Observed Ageing in FT0 MCP-PMTs}
\begin{figure}[!ht]
\centering
\begin{subfigure}{.24\textwidth}
  \centering
  \includegraphics[width=.9\linewidth]{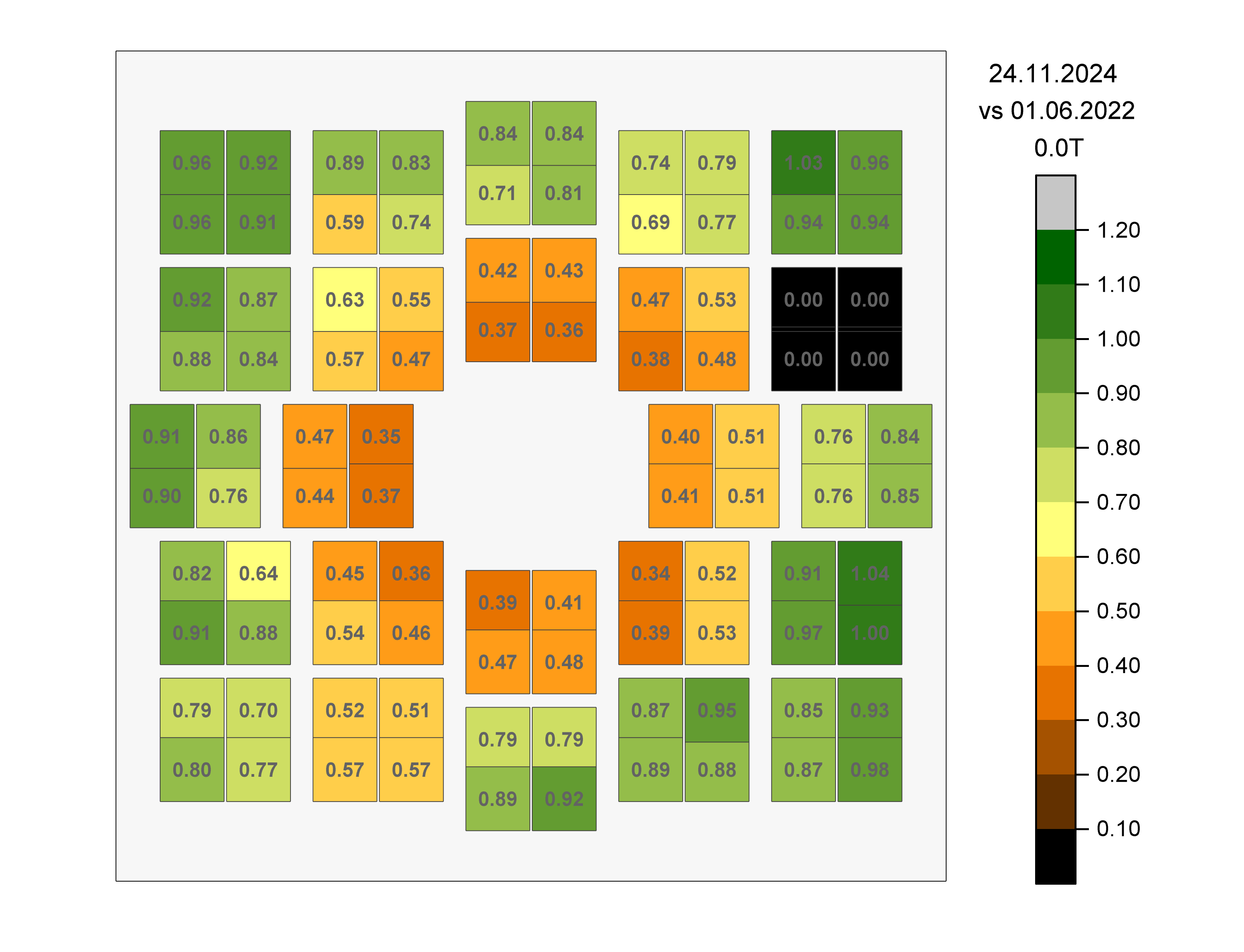}
  \caption{FT0-A}
  \label{fig:ageing_FT0-A}
\end{subfigure}%
\begin{subfigure}{.24\textwidth}
  \centering
  \includegraphics[width=.9\linewidth]{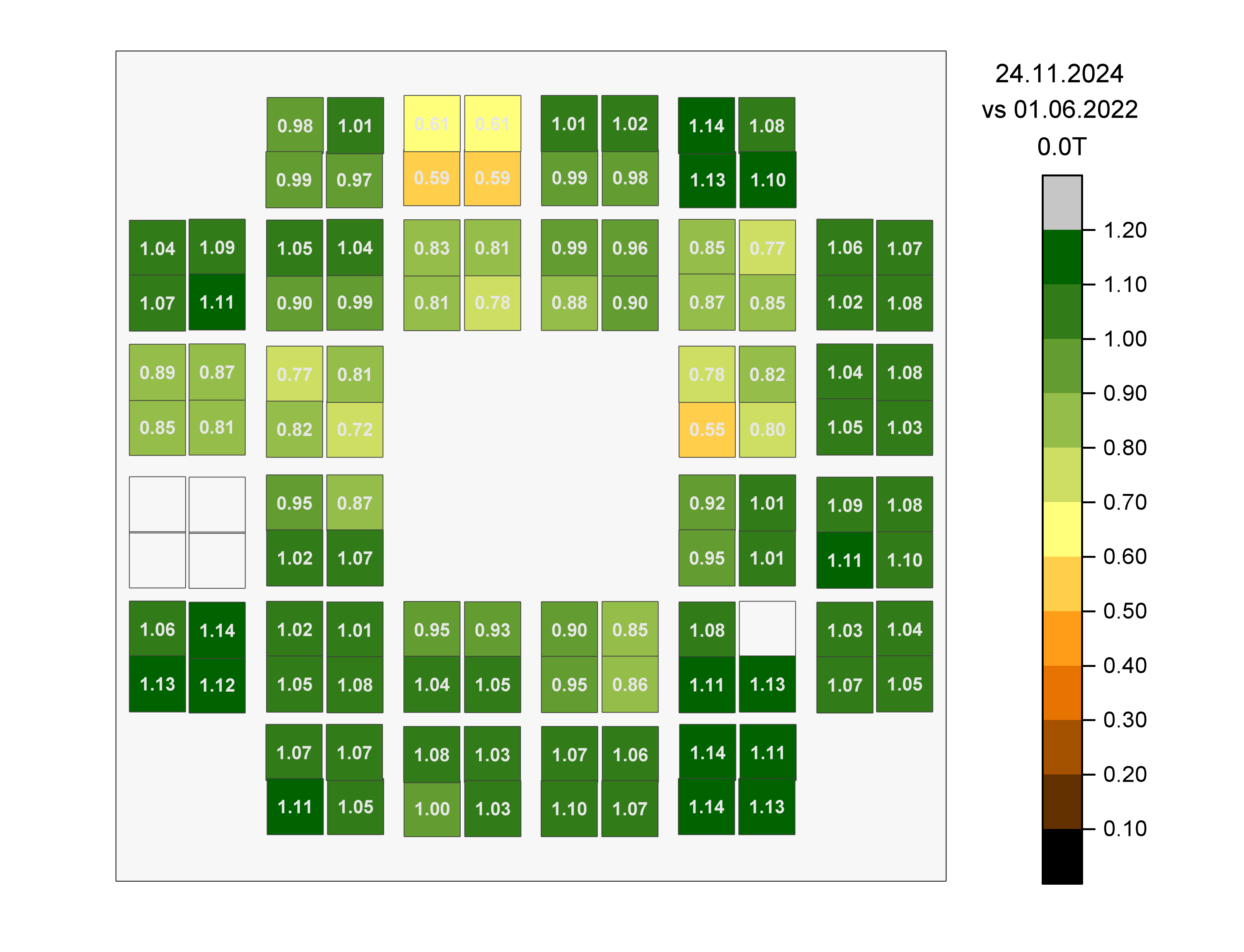}
  \caption{FT0-C}
  \label{fig:ageing_FT0-C}
\end{subfigure}
\caption{Measured signal ratio in November 2024 to a reference from June 2022, indicating gradual gain reduction in some FT0 MCP-PMTs.}
\label{fig:test}
\end{figure}

With higher-than-anticipated integrated luminosities, certain channels in the FT0 arrays show reduced gain. Figure~\ref{fig:test} demonstrates the ratio of signals from MCP-PMTs in November 2024 relative to a June 2022 reference. Although performance remains acceptable for Run~3, proactive measures are being taken to ensure long-term stability.  With gain compensation using high voltage, current performance remains within acceptable thresholds for timing and trigger requirements, however there are plans to replace the most affected 11 sensors during the LS3. These preventative measures aim to ensure the long-term stability and precision of the FT0 subsystem, particularly as higher luminosities in upcoming LHC runs could accelerate the ageing effects further.

\subsection{Front-End Electronics Roadmap}

Although the first-generation FEE has successfully handled Run~3 demands, it is being redesigned for improved signal processing and reliability. The key goals include:
\begin{itemize}
    \item Larger dynamic range in amplitude measurements, essential for scintillation signals from the FV0 and FDD under increased particle flux.
    \item Improve CFD response for high amplitude signals.
    \item Reduce high-frequency noise modifying CFD time distributions for low amplitude signals.
    \item Reduce the effect of oscillations on signal trailing edge on the CFD performance.
    \item Online tagging of pileup and background events (signal shape analysis).
\end{itemize}

\begin{figure}[!ht]
    \centering
    \includegraphics[width=.7\linewidth]{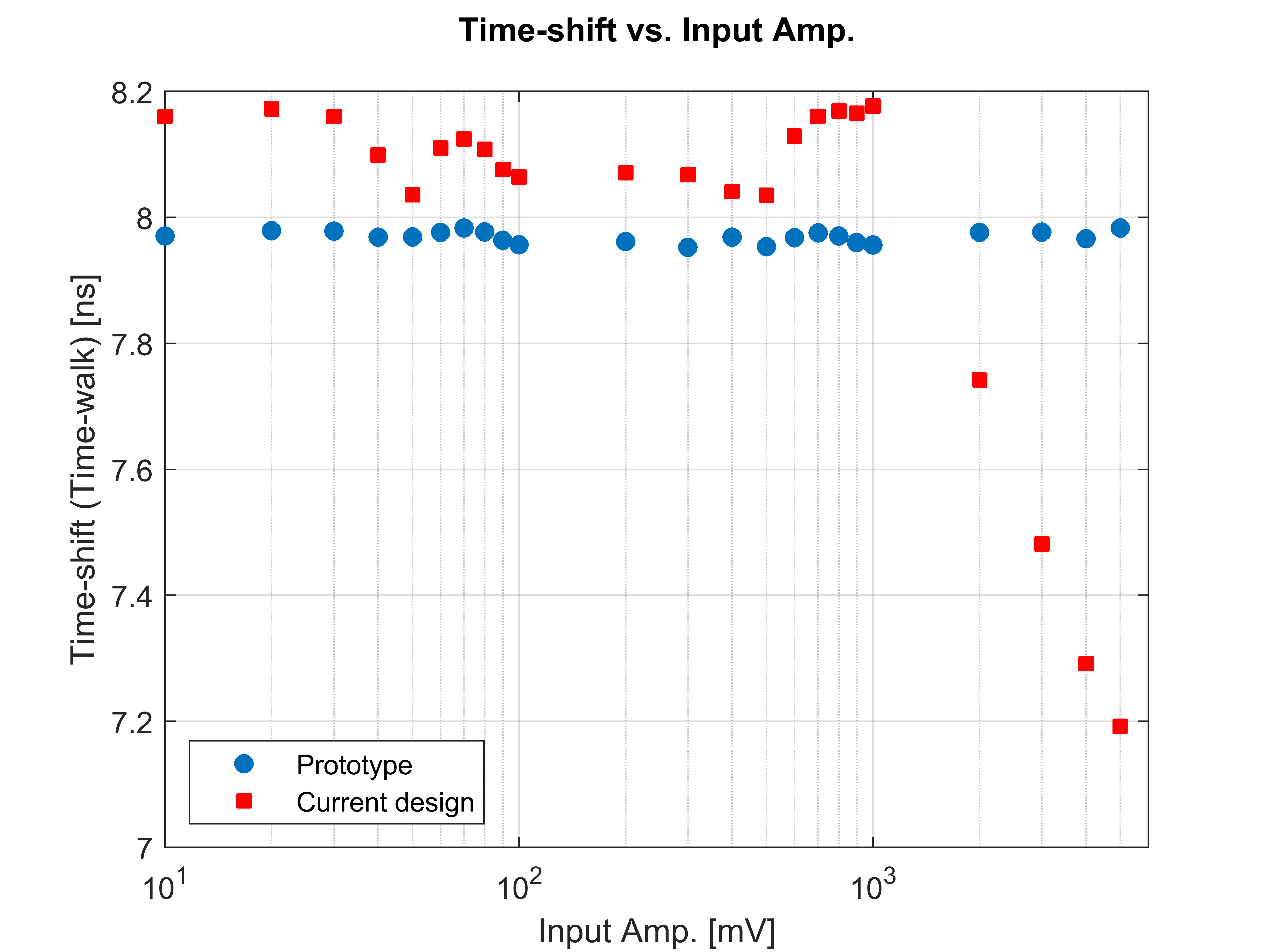}
    \caption{Time-walk as a function of input amplitude for the current design and the prototype.}
    \label{fig:CFD}
\end{figure}

The enhancements in the FEE are concentrated around advancing measurement capabilities and data processing. The modified PM (Processing Module) in version 1 that we are planing to test in March 2026, will support dual energy measurements and can be equipped with either the existing or an improved overlay, potentially integrating a new integrator. After the LS3 we plan to test version 2 that transitions towards a fully digital readout path, including a digital integrator and a new Time-to-Digital Converter (TDC) that could leverage either MGT FPGA technology or a bespoke ASIC developed in Poland. Now we are testing the CFD prototype which indicates a significant improvement in the pulse response time with minimal sensitivity to input amplitude, achieving values close to the theoretical 7.35 ns. Figure~\ref{fig:CFD} demonstrates the time-walk as a function of input amplitude for the current design and the prototype.

\subsection{Integration of ALFRED with the FIT DCS}

Unlike most ALICE subdetectors, which send control commands and retrieve status data via GBT, the FIT setup communicates over IPbus. IPbus provides transaction-oriented data exchanges—read and write operations—over standard TCP/IP networks. While this affords flexibility and scalability, it also differs from the GBT-based approach that ALFRED typically supports.

\begin{figure}[!ht]
    \centering
    \includegraphics[width=.9\linewidth]{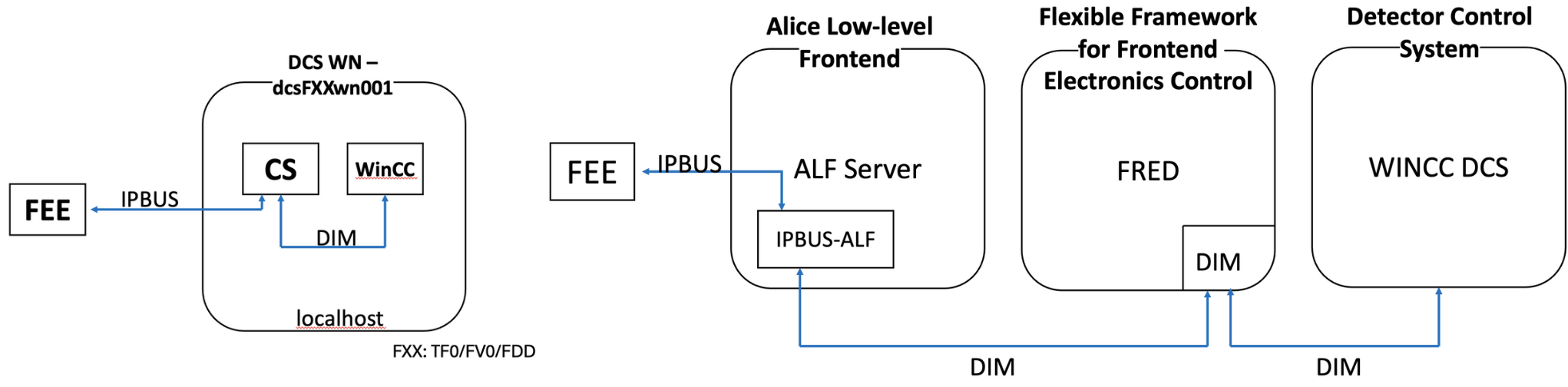}
    \caption{Planned upgrade of the Detector Control System for FIT, employing IPbus for FEE communication and the ALFRED layer for uniform integration.}
    \label{fig:DCS_Upgrade}
\end{figure}

To incorporate IPbus, an IPbus-ALF service was developed \cite{Roslon:2025tym}. Rather than running on the First Level Processor (FLP, as is done for GBT-based systems), this IPbus-ALF program is installed on a separate machine that shares a secure local network connection with the FIT front-end modules. Figure~\ref{fig:DCS_Upgrade} shows the schematic of the data path in the current version and the one implemented in the FV0 detector.

Through this chain, standard ALFRED functionalities --- such as configuration management, error handling, and run control --- become available to the IPbus-based electronics without the need to rewrite existing firmware. The IPbus-ALF server essentially replaces the CRU-based GBT access with IPbus transactions, preserving the rest of the ALFRED architecture.

\section{New Forward Detector Concepts Beyond Run 3}
\label{sec:newFD}

\begin{figure}[!ht]
    \centering
    \includegraphics[width=.9\linewidth]{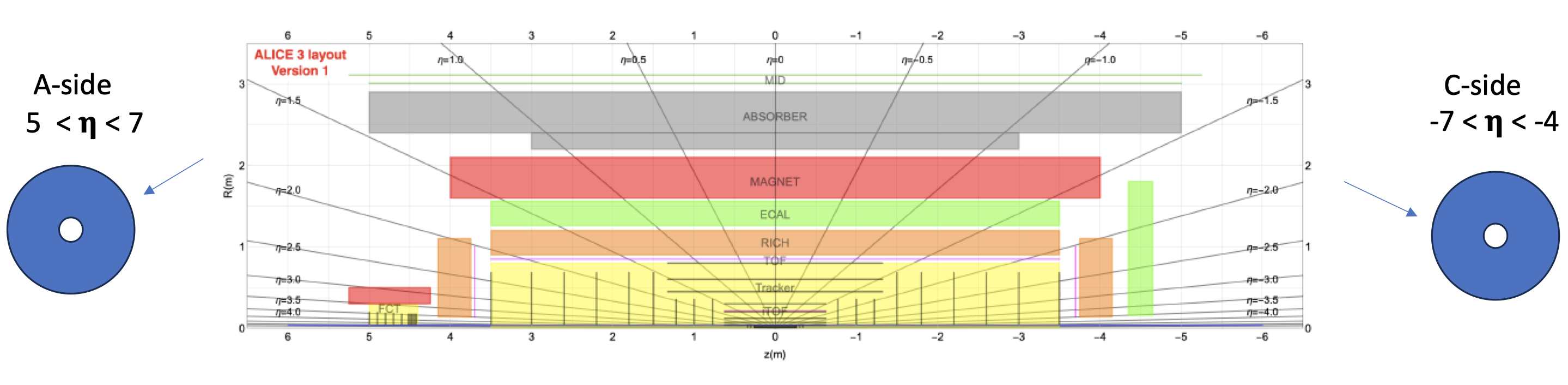}
    \caption{Concept for a new Forward Detector in the next-generation ALICE3 apparatus, potentially replacing or augmenting the current FDD.}
    \label{fig:FIT_Upgrade}
\end{figure}

Figure~\ref{fig:FIT_Upgrade} shows a conceptual design for a new Forward Detector (FD) which is under consideration for the ALICE 3 upgrade, designed to cover the very-forward pseudorapidity range from approximately \(\eta \approx 4\) to \(\eta \approx 7\) on the A-side, and from \(\eta \approx -7\) to \(\eta \approx -4\) on the C-side. The detector would be positioned near the current FDD location.

Conceptually, the FD builds upon the existing FIT--FV0 approach by utilising segmented organic scintillator disks (e.g.\ EJ204, EJ208, or radiation-hard variants such as PEN/PET) coupled via optical fibres to fast photomultiplier tubes (PMTs) or silicon photomultipliers (SiPMs). These components will provide both time and charge measurements, thereby facilitating efficient event triggering and characterisation at forward rapidities.

Key R\&D efforts focus on ensuring radiation hardness at the Mrad level, reducing latencies below 25 ns for bunch-crossing rates near 24 MHz, and maintaining a sufficiently large dynamic range to accommodate heavy-ion collisions of varying centralities.

In parallel, the existing knowledge from the FT0, FV0, and FDD upgrades forms a testbed for these future designs, guiding sensor material studies and electronics architecture.

\section{Summary and Outlook}
\label{sec:summary}

The Fast Interaction Trigger system has become a linchpin of ALICE's data-taking: it delivers prompt minimum-bias triggers, assists in luminosity and background monitoring, and provides collision-time references critical for TOF-based particle identification. Run~3 operations have confirmed its reliability, with time resolutions in the range of 17\,ps for proton--proton and 4.4\,ps for Pb--Pb collisions.

Ageing of certain MCP-PMT channels and the move to higher luminosities have motivated a need of sensor exchanging. New front-end boards, featuring enhanced dynamic range and improved noise handling, are under development. In parallel, the ALICE Low-Level Front-End Device framework will standardise communication between the DCS and the electronics. Together, these changes ensure the system remains optimised through the end of Run~4.

Beyond these mid-term upgrades, a Forward Detector concept is being explored to extend the acceptance to pseudorapidities up to about 7. Integrating the lessons learned from the current FIT configuration, such a forward system could enrich the ALICE’s physics scope in future high-luminosity collisions, enabling precision global measurements in heavy-ion and proton--nucleus data sets. Future R\&D will determine whether scintillators, or novel composite materials prove most suitable for the harsh forward environment.

In summary, the FIT continues to evolve to meet ALICE’s stringent requirements. The newly presented improvements, particularly in the FEE and the DCS integration (ALFRED), demonstrate the collaboration’s commitment to sustaining world-class performance and enabling innovative detection concepts.


\section{Acknowledgments}
This work was supported by the Polish Ministry of Science and Higher Education under agreements no. 5452/CERN/2023/0, 2022/WK/01, 2023/WK/07 and ``The Excellence Initiative - Research University'' programme.

\bibliography{mybibfile}

\end{document}